\documentclass{aa}
\usepackage{graphicx}

\begin{document}

\sloppypar

   \title{RXTE all-sky slew survey. Catalog of X-ray sources at $|b|>10^\circ$}

   \author{M. Revnivtsev\inst{1,2}, S. Sazonov\inst{1,2},
K. Jahoda\inst{3}, M. Gilfanov \inst{1,2}}

   \offprints{mikej@mpa-garching.mpg.de}

   \institute{Max-Planck-Institut f\"ur Astrophysik,
              Karl-Schwarzschild-Str. 1, D-85740 Garching bei M\"unchen,
              Germany
        \and   
              Space Research Institute, Russian Academy of Sciences,
              Profsoyuznaya 84/32, 117997 Moscow, Russia
        \and 
                Laboratory for High Energy Astrophysics, Code 662,
              Goddard Space Flight Center, Greenbelt, MD 20771, USA
            }
  \date{}

        \authorrunning{Revnivtsev et al.}
        \titlerunning{}
 
  \abstract{We report results of a serendipitous hard X-ray (3--20
keV), nearly all-sky ($|b|>10^\circ$) survey based on RXTE/PCA
observations performed during satellite reorientations in
1996--2002. The survey is 80\% (90\%) complete to a 4$\sigma$ limiting flux
of $\approx 1.8$ (2.5) $\times 10^{-11}$ erg s$^{-1}$ cm$^{-2}$ in the
3--20~keV band. The achieved sensitivity in the 3--8~keV and 8--20~keV
subbands is similar to and an order of magnitude higher than that of
the previously record HEAO-1 A1 and HEAO-1 A4 all-sky surveys,
respectively. A combined $7\times 10^3$~sq.~deg area of the sky is
sampled to flux levels below $10^{-11}$ erg s$^{-1}$ cm$^{-2}$ (3--20
keV). In total 294 sources are detected and localized to better than
1~deg. 236 (80\%) of these can be confidently associated with a known
astrophysical object; another 22 likely result from the superposition
of 2 or 3 closely located known sources. 35 detected sources remain
unidentified, although for 12 of these we report a likely soft X-ray
counterpart from the ROSAT all-sky survey bright source catalog. Of
the reliably identified sources, 63 have local origin (Milky Way, LMC
or SMC), 64 are clusters of galaxies and 100 are active galactic
nuclei (AGN). The fact that the unidentified X-ray sources have hard
spectra suggests that the majority of them are AGN, including highly
obscured ones ($N_{\rm H}>10^{23}$~cm$^{-2}$). For the first time we
present a $\log N$--$\log S$ diagram for extragalactic sources above
$4\times 10^{-12}$ erg s$^{-1}$ cm$^{-2}$ at 8-20 keV. 
   }

   \maketitle

%

\section{Introduction}

The deep surveys performed recently in the standard X-ray band (2--10~keV)
with the Chandra and XMM-Newton observatories (e.g. \cite{chandra},
\cite{xmm}) have convincingly proved the extragalactic origin of the
cosmic X-ray background (CXB). Hundreds of point sources detected in these
surveys provide us with a wealth of information about the distant
Universe. However, due to the very small sky coverage of these
surveys, they are practically unsuitable for the study of the local
Universe ($z\la 0.3$). Medium-sensitivity
($10^{-13}$--$10^{-12}$~erg~cm$^{-2}$~s$^{-1}$) X-ray surveys, such as
those performed with ASCA (\cite{gis1}) and BeppoSAX
(\cite{giommi2000}), cover larger areas of the sky ($\la
10^2$~sq. deg) but also cannot sample efficiently the Universe within 
$\sim 500$~Mpc of us. In this regard, the results of the soft X-ray
($<2$~keV) all-sky survey carried out with the ROSAT observatory
(e.g. \cite{rbsc}) are extremely important, but these cannot be
directly extrapolated into the $>2$~keV energy band. Therefore, our
knowledge of the statistical properties of the local population of
hard X-ray sources still rests largely on the snapshot of the whole sky
taken in the 2--100~keV energy band more than 20 years ago by the
different experiments on board the HEAO-1 observatory, A1 (\cite{a1}),
A2 (\cite{a2}) and A4 (\cite{a4}). It is only now that we have the
possibility to undertake a new hard X-ray (3--20~keV) all-sky survey
at similar (below 10~keV) and much better (above 10~keV) sensitivity
provided by the RXTE observatory.

The Rossi X-ray Timing Explorer (RXTE, \cite{rxte}) was launched at the end 
of 1995 and has now been successfully operating for more than 7
years. The mission was primarily designed to study the variability of
X-ray sources on time scales from sub-milliseconds to years
(e.g. \cite{swank_ns}). The maneuvering capability of the
satellite combined with the high photon throughput of its main
detector (PCA) has also made it possible to carry out a series of 
Galactic Bulge scans aimed at the detection of new transient sources and
following the long-term behavior of known X-ray sources
(\cite{craig00}). In addition, over its still continuing life time,
RXTE/PCA has collected a large amount of data of slew observations
covering almost the entire sky. In the current work we use these data
to perform an all-sky survey in the 3--20 keV energy band. The
relatively narrow field of view of the PCA instrument allows us to
localize sources to better than 1 deg and thus effectively avoid
source confusion after we restrict our consideration to Galactic
latitudes $|b|>10^\circ$.

We note that the RXTE/PCA slew data have previously been utilized to 
reconstruct the average spectrum of the CXB (\cite{cxbpaper}). 

\section{Data analysis}

\subsection{Data selection and reduction}

One of the main instruments aboard the RXTE observatory  is the
Proportional Counter Array (PCA), an X-ray spectrometer. It consists of 5
nearly identical Proportional Counter Units (PCUs). Each
PCU is sensitive to photons with energies 2--60 keV, reaching the
maximum effective area at $\sim$ 7 keV. For an X-ray source with a Crab-like spectrum, most of the counts would be detected at energies below 10
keV. However, a significant effective area, 300--500 cm$^2$ per PCU,
is also available in the 10--20 keV energy band. 

\begin{figure}
\includegraphics[width=\columnwidth]{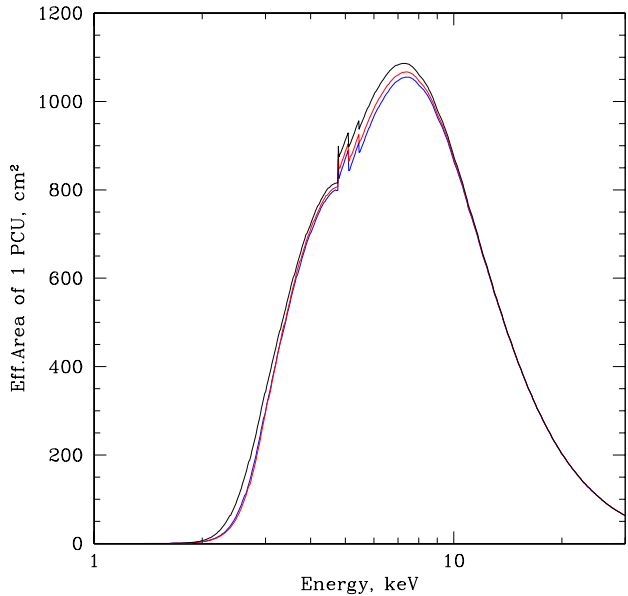}
\caption{Effect of high voltage variations on the PCU effective
area. Shown is the effective area of PCU2 on Apr. 17,1996 (epoch
3), Dec. 06,1999 (epoch 4) and June 15,2002 (epoch 5) \label{areas}}
\end{figure}

\begin{figure*}[t]
\includegraphics[width=\textwidth,bb=50 250 560 528,clip]{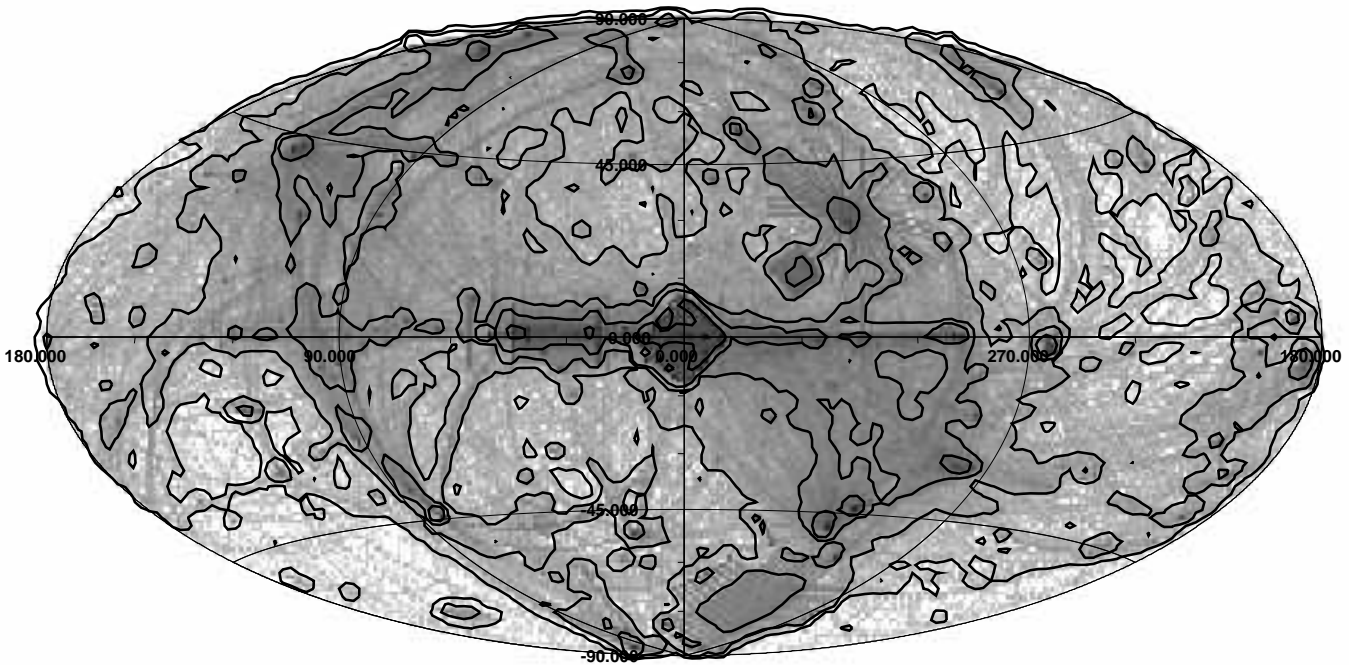}  
\caption{Sky exposure map (in Galactic coordinates) of the RXTE/PCA slew
survey performed during April 1996--July 2002, recalculated per one
PCU area. Smoothed contours are drawn at 100, 300, 1000, 3000, 10000
and 30000 s~deg$^{-2}$.  
\label{exposuremap}} 
\end{figure*}

The slewing rate of the RXTE satellite between targets is $<$0.1
deg~s$^{-1}$. One can make use of data collected during such reorientations to 
build maps of the sky. We have utilized the slew parts of all RXTE
observations performed from April 15, 1996--July 16, 2002. This
amounts in total to approximately 50,000 observations of typical
length 200--500~s. The observational period before April 15, 1996
(High Voltage Epochs 1 and 2) was 
excluded from the analysis because during that time the PCA had
significantly different gain and dependence of the effective area on
energy. Since these early observations constitute only $\sim 2.5$\% of
all available slew/scan data, their rejection has
negligible effect on the net exposure time of our survey, which is 20.2 Ms.

The data reduction was done using standard tools of the
LHEASOFT/FTOOLS 5.2 package. In this version of the software the effective area
of the PCA detectors is slightly ($\sim$~11--12\%) underestimated,
which leads to an overestimation of the fluxes of X-ray sources 
(see \cite{cxbpaper}, Jahoda et al. 2004, in preparation). We have
made a correction for this factor in our analysis.

For the background modeling we used the faint source (''L7\_240'') CM model
(http://heasarc.gsfc.nasa.gov/docs/xte/recipes/ pcabackest.html). 
The background model includes by design both the CXB and instrumental
background. Therefore, the background subtracted rate for ''blank sky''
observations is expected to be consistent with zero within 
spacial fluctuations of the CXB. The rms amplitude of these
fluctuations (cosmic variance) was earlier estimated for RXTE/PCA to be 
$\sim 7$\%, or $\sim 1.5\times 10^{-12}$ erg~s$^{-1}$~cm$^{-2}$, in 
the 2--10 keV energy band (\cite{cm_model}, \cite{cxbpaper}).

At low galactic latitude, there is diffuse X-ray emission from the Galactic
ridge and bulge. Using results of \cite{iwan82} and
\cite{mikej_diffuse}, we estimate that this component does not exceed
$\sim 2-3\times 10^{-12}$ ergs~s$^{-1}$~cm$^{-2}$ at $|b|>10^\circ$ in
the central part of the Galaxy and continues to decrease with distance
from the Galactic center (in $l$ and $b$). We therefore neglect the
Galactic diffuse emission in our subsequent analysis.

In order to reduce the effect of the uncertainties in the PCA
background subtraction and increase our sensitivity to source 
detection, we used only the data of the first layer (LR1) for all
PCUs. The effective area of a PCU (the first upper anode layer only)
is shown as a function of photon energy in Fig. \ref{areas}. Different
curves in the figure represent different high voltage epochs of the
PCA. The observed variations of the effective area prove to be mainly
due to an increasing amount of xenon in the propane layer (Jahoda et
al. 2004, in preparation) rather than due to the change of the high voltage.

We utilized data of Standard2 mode of the PCA, which are available for all
observations. Standard2 mode provides the possibility of energy band
selection and also 16~s time resolution. Energy band selection is very
important, because by using a band narrower than the PCA total
energy range (2--60 keV) we can significantly reduce the instrumental
background and therefore strongly increase our sensitivity to faint
sources. However, during a 16-s interval the center of the PCA field of view
can move up to 1--1.5$^\circ$, which limits the spatial resolution of
maps constructed from Standard2 data (see \S\ref{map} for further
discussion of this issue).

Data were filtered by the following criteria. We filtered out
Earth occulted data (by applying the criterion $ELV>10$). Our study of
the PCA background subtraction uncertainties and their dependence on
the $ELECTRON$ rate demonstrated that for our purposes (weak source
detection) it is better to use the criterion
$ELECTRON\_PCU0,1,2,3,4<0.085$ than the standard condition $<0.1$. In
addition, we excluded all data from PCU0 obtained later than May 12,
2000, because the loss of the upper propane veto layer of this PCU has
resulted in strongly increased background subtraction
uncertainties. After applying the above filters we are left with 9.3 
Ms exposure time worth of good data, covering almost all of the
sky. Since during some slews not all 5 PCUs were operational, the resulting
exposure time recalculated per one PCU is 26.9 Ms. The corresponding
exposure map is shown in Fig. \ref{exposuremap}. 

\subsection{Sensitivity and completeness of the survey}

To avoid dealing with substantial source confusion due to high density
of Galactic X-ray sources, we excluded observations performed at 
low Galactic latitude $|b|<10^\circ$ from the construction of a catalog
of detected sources (see \S\ref{cat}). The total accumulated
exposure time per PCU corresponding to the $|b|>10^\circ$ region (with
the total area of 34,090~sq. deg) is 16.6~Ms. 

\begin{figure}
\includegraphics[width=\columnwidth]{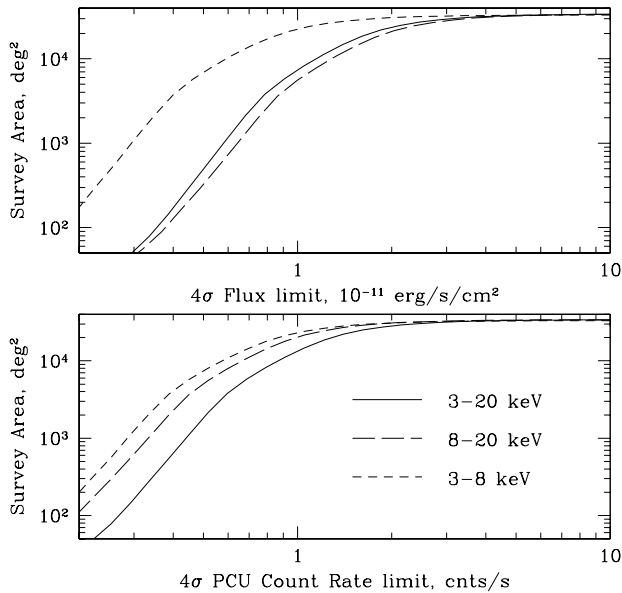}
\caption{
Sky coverage at $|b|>10^\circ$ of the RXTE/PCA slew survey as a
function of the minimum count rate (lower panel) and energy flux
(assuming a Crab-like source spectrum, upper panel) needed for detection of
a source at 4$\sigma$ confidence level in different energy bands.
\label{exposure_area}} 
\end{figure}

Fig. \ref{exposure_area} shows the effective survey
area as a function of the minimum PCU count rate allowing 4$\sigma$ source
detection in the 3--20~keV, 3--8~keV and 8--20~keV bands, and also as
a function of an energy flux threshold for 4$\sigma$ detection of
sources with a Crab-like spectrum (power law with a photon index
$\Gamma=2$). The survey is 80\% (90\%) complete at $|b|>10^\circ$ down
to 3--20~keV, 3--8~keV and 8--20~keV count rates of 1.8 (2.5), 1.2
(1.6) and 1.3 (1.8) cnt~s$^{-1}$. For $\Gamma=2$ sources, these
numbers correspond to the following limiting fluxes: 2.3 (3.3), 1.2
(1.6) and 2.5 (3.4) $\times 10^{-11}$~erg~s$^{-1}$~cm$^{-2}$. The
sensitivity of the RXTE slew survey below 10~keV is thus comparable
to/slightly higher than that of the HEAO-1 A1/A2 surveys
(\cite{a1,a2}). In the 10--20~keV energy band, the current 
slew survey has record sensitivity for all-sky hard X-ray surveys,
being an order of magnitude deeper than the HEAO-1/A4 survey
(\cite{a4}). 

\subsection{Construction of the all-sky map; detection and
localization of sources}
\label{map}

Relatively fast and irregular slews of the RXTE satellite do not allow
us to detect weak sources using the procedures previously applied to
the analysis of the UHURU and HEAO-1 all-sky surveys (\cite{4u}, \cite{a2},
\cite{a1}). For this reason, our source detection algorithm is based
on analyzing averaged maps of the sky instead of using individual
scans. 

The accuracy of source localization depends on the source brightness
and variability as well as on the RXTE slewing speed, but is always better than
$1-1.5^\circ$ for a single slew across a source. The localization of a
source that has been scanned across several times can be significantly
improved by fitting its observed brightness profile on the sky with a
model of the collimator response. The locations of extremely stable
and strong sources such as the Crab nebula or brightest clusters
of galaxies can thus be recovered within few arcminutes.

A sequence of several steps, as described below, was performed to build
the all-sky map and to detect and localize sources in the sky. First, a 
map of the sky for each PCU was obtained by ascribing the measured,
background subtracted count rate to the celestial position toward
which the center of the PCA field of view was pointed at the middle of
each 16-s bin of data. Such maps were produced in the total
3--20~keV energy band as well as in the subbands 3--8~keV and
8--20~keV. Then, exposure-weighted averaging of the maps from the
individual PCUs was done to obtain raw maps of the sky in the three
energy bands.

\begin{figure}
\includegraphics[width=\columnwidth]{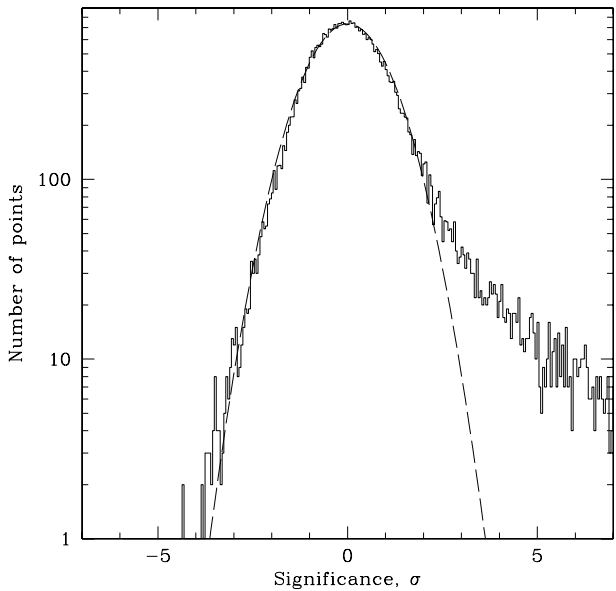}
\caption{Distribution of signal-to-noise ratios for $\sim 34,000$
$1^\circ\times 1^\circ$ bins making up the $|b|>10^\circ$ sky map in
the 3--20~keV band. The dashed line is a Gaussian with zero
mean and unity dispersion, expected for pure statistical noise.
\label{sigpoints}}
\end{figure}

The derived all-sky maps are of very high statistical quality as
demonstrated in Fig. \ref{sigpoints}, where we present the distribution of the 
observed signal-to-noise ($S/N$ \footnote{Here $N$ means purely
statistical noise of the measurements. We do not include here the 
variations of the CXB brightness level, which are smaller than the
statistical uncertainty of a source flux measurement for typical exposure times
$\sim$100 s})  ratios for the $\sim 34,000$ $1^\circ\times
1^\circ$ bins making up the sky map at $|b|>10^\circ$ in the 3--20~keV
band. The chosen pixel size assures that neighbouring
bins are practically independent, because the effective field of view
of RXTE/PCA is 1 sq. deg. In the case of pure statistical noise, the
$S/N$ distribution is expected to be a Gaussian with zero mean and
unity dispersion, which is quite similar to what is actually observed
for $S/N< 2$ (see Fig. \ref{sigpoints}). Given this excellent
agreement, we require that a source have a $S/N>4$ in the 3--20~keV
band to be considered detected. With this criterion, we allow the detection of
approximately one fake source over the whole sky at
$|b|>10^\circ$. 

\begin{figure*}[t]
\includegraphics[width=\textwidth]{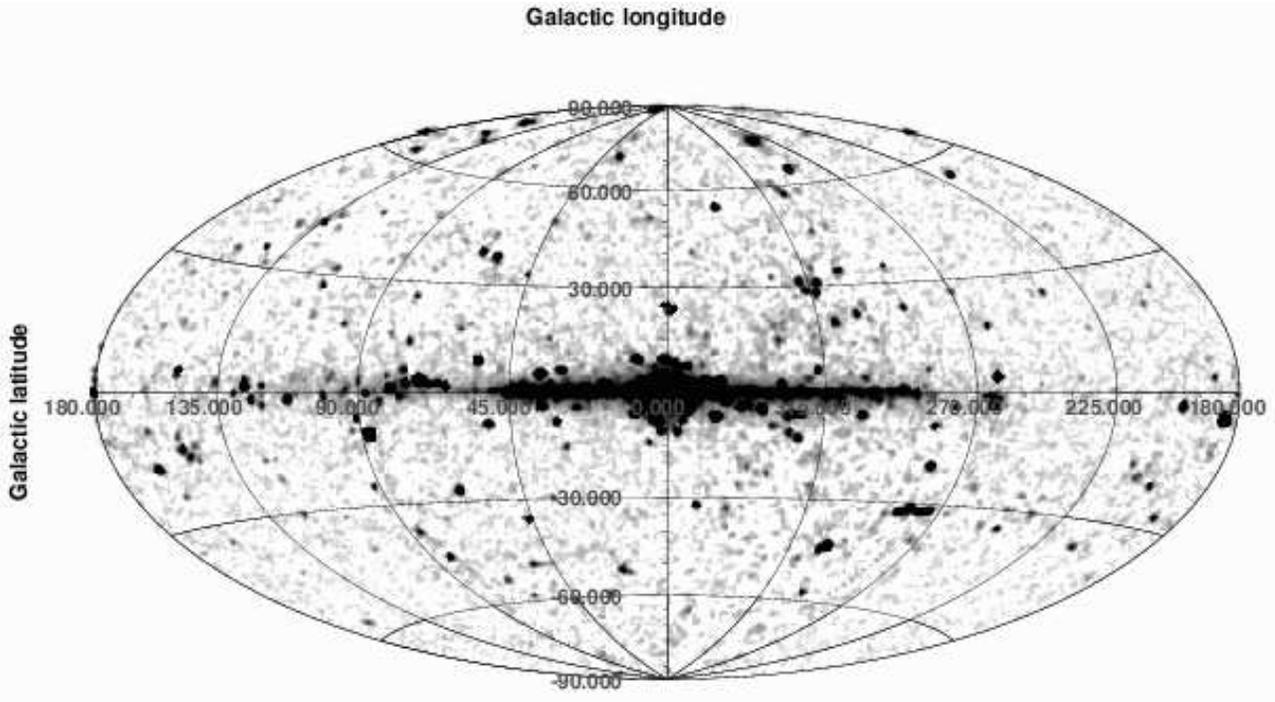} 
\caption{All-sky map obtained with RXTE/PCA in the 3--20
keV energy band \label{skymap}}
\end{figure*}

A simply binned raw map of the sky does not provide us with the best signal to
noise ratio for the search of point sources. In order to maximally use
the available statistics we convolved the raw all-sky maps with the
response of the RXTE/PCA collimator (\cite{jahoda96}). The resulting
3--20~keV all-sky map is shown in Fig.~\ref{skymap}. This map was used
to detect sources by searching for peaks characterized by a $S/N$
ratio of more than 4. The positions and fluxes of so found sources
were subsequently determined from a $\chi^2$ fit of the PCA collimator 
model to the raw all-sky map. In our catalog of detected sources (see
\S\ref{cat}), their best-fit positions are given with a 1$\sigma$
statistical uncertainty. 

However, the above procedure is not optimal for detection and
localization of sources covered by only a few RXTE slews and
for strongly variable sources. We therefore also searched the
1$^\circ$-binned raw all-sky map for pixels that have $S/N>4$. This
has allowed us to detect several addional sources, missed in the
previous step (convolved sky analysis). For these sources, a
conservative estimate of the total localization uncertainty is
provided in the catalog (see \S\ref{cat}). In all cases, the error is
less than or equal to $1^\circ$.

\subsection{Flux measurement and source confusion}
\label{flux}

Measurement of source fluxes in our survey is subject to some systematic
uncertainties. Some uncertainty may arise from insufficiently
accurate determination of source positions. The further the inferred position
is from the actual one, the smaller flux can be ascribed to the source
because of the response of the collimator falling off with increasing
offset from the optical axis. This should not however result in a biased 
estimate of the flux of a source that has been covered by numerous
slews, because in that case both the source position and its flux are
determined simultaneously from a fit of the model collimator response
to the source image. However, the effect may be important for poorly
covered sources and also for variable and transient sources. As most
of the sky in our survey is covered by more than 5 slews, the effect
is expected to be generally small. 

We could estimate the accuracy of the flux determination if we had a sample of 
persistent sources with known X-ray fluxes. Fortunately, there are
64 clusters of galaxies detected in our survey (see \S\ref{cat}) for
most of which we can compare their measured RXTE/PCA 3--8 keV count  
rates with the corresponding values evaluated from the fluxes and gas
temperatures derived with ROSAT and ASCA observatories (\cite{max},
\cite{ebeling}, \cite{ikebe}). The fact that clusters of galaxies are
not point sources does not strongy affect our flux estimates because
of the low  ($\sim 1^\circ$) angular resolution of our
measurements. The strongest deviations from the true source flux, on
the order of 20-30\%, may arise in the case of largest clusters of galaxies
such as the Coma.

\begin{figure}
\vbox{
\includegraphics[bb=40 188 584 592,clip,width=\columnwidth]{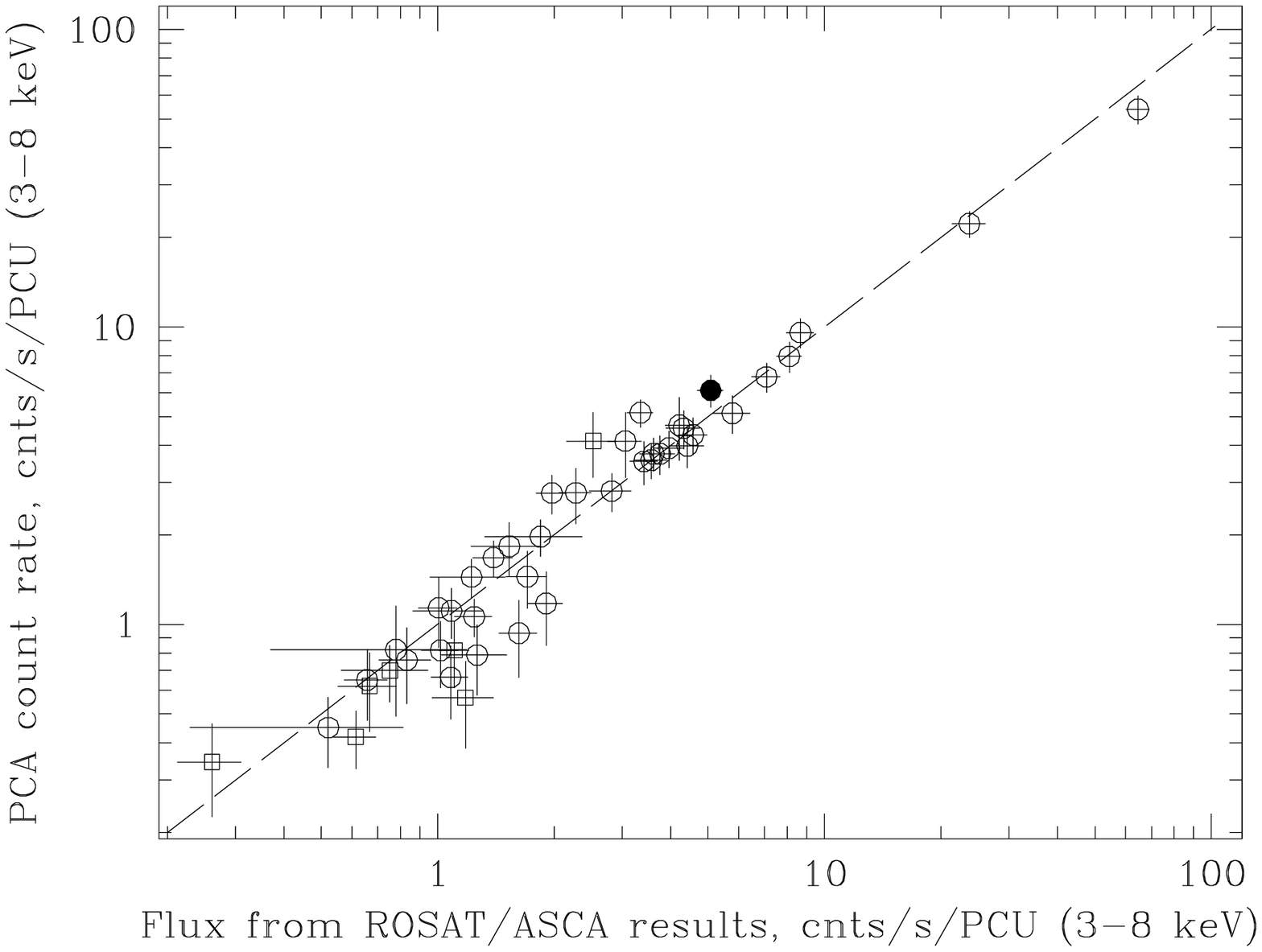}
\includegraphics[bb=50 188 584 416,clip,width=\columnwidth]{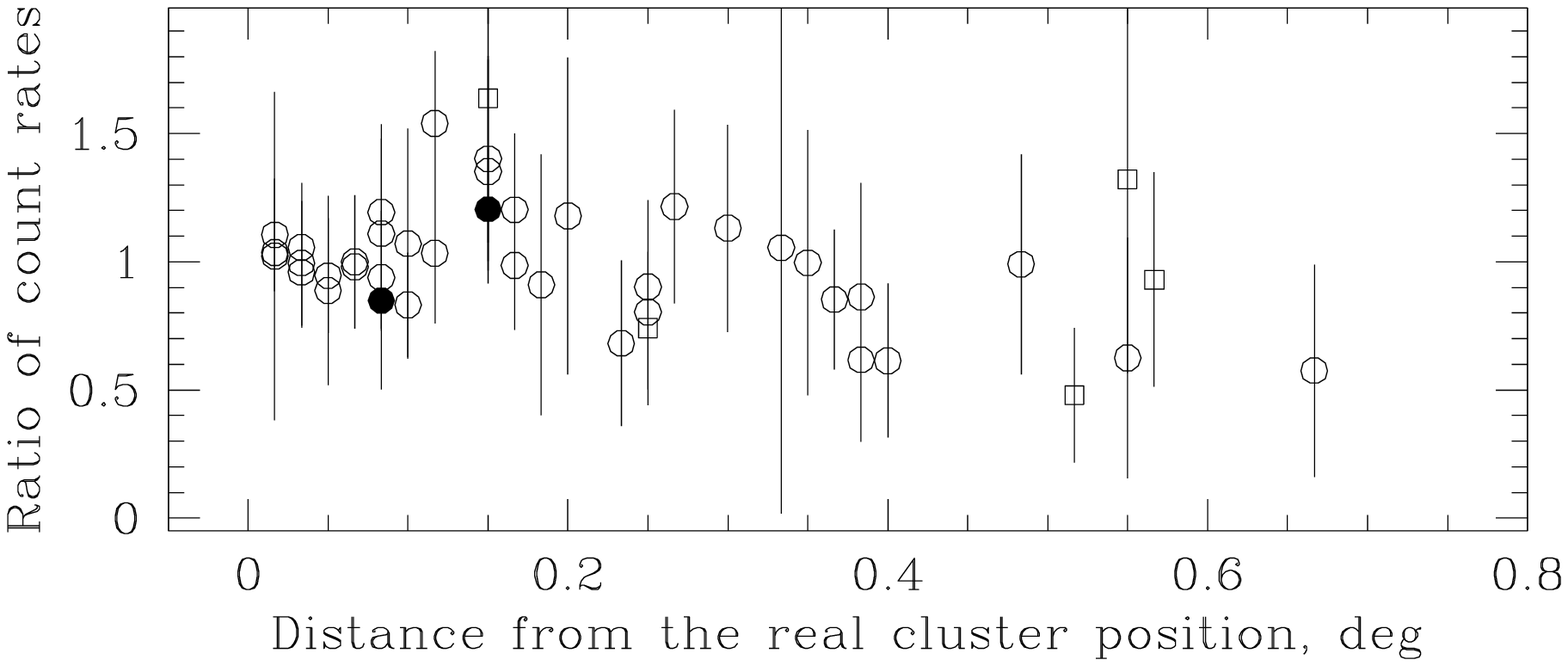}
}
\caption{{\bf Upper panel}: Count rates measured in the RXTE/PCA slew
survey vs. count rates predicted from fluxes and gas temperatures
measured with ROSAT and ASCA for 45 clusters of galaxies. The data
are taken from Ikebe et al. (2002) -- open circles, Markevitch (1998) --
solid circles, and Ebeling et al. (2000) -- open squares. {\bf Lower
panel}: Ratio of the measured and predicted count rates as a function of
distance of the localization from the true position of clusters.
\label{fluxes}}
\end{figure}

 It can be seen from Fig. \ref{fluxes}
that the agreement between the two sets of count rates is very good and
there seems to be no significant bias in our flux determination even
for angular offsets as large as $40^\prime$. Since for the absolute
majority of sources in our catalog that can be confidently
associated with a known astrophysical object the  RXTE location is
within $40^\prime$ of the true position, we will not attempt to
correct any of the measured fluxes for the possible bias described
above. We note however that the fluxes of the (weak and poorly
localized) unidentified sources in our catalog may be systematically
underestimated.

Because of the fairly large field of view of the PCA ($\sim
1$~sq. deg), source confusion presents some problem for  
our survey. Confusion may manifest itself through the appearance of two or
more sufficiently bright sources within our beam and lead to
ambiguous source identification and uncertainty in source flux determination.
Using cumulative source counts interpolated from HEAO-1/A2, ASCA and
BeppoSAX measurements, or the consistent estimate from the current survey
(see \S\ref{nflux}), and taking into account the sky coverage 
and sensitivity of our survey (Fig. \ref{exposure_area}), we may estimate
that there can be of the order of 10 unresolvable pairs and 
larger groups of sources, detectable as single sources with $S/N>4$,
on the sky at $|b|>10^\circ$, assuming a uniform
distribution of sources. We do find a comparable number of confusion
cases (groups of known X-ray sources) in our catalog (see \S\ref{cat}).

The uncertainty of the flux determination can be expressed in terms of
confusion variance $\sigma_{\rm conf}$. In the case of a power-law form of
cumulative source counts (the number of sources with a flux greater
than $f$ per unit solid angle), $N(>f)=C f^{-\alpha}$, the confusion noise is
given by  
$$
f/\sigma_{\rm conf}(f)=\left({\alpha\over{2-\alpha}}\Omega N(>f)\right)^{-1/2},
$$ 
where $\Omega$ is the effective solid angle of the instrument beam 
(e.g. \cite{hackhou87}). For the standard value $\alpha=1.5$ and 
the observed density of sources, we find that confusion sets in at count
rates below $4\sigma_{\rm conf}\approx 0.5$~cnt~s$^{-1}$ in the
3--20~keV band, or equivalently (assuming a Crab-like spectrum) at
fluxes below $7\times 10^{-12}$~erg~s~cm$^{-2}$.

Summarizing the above discussion, we can conclude that our estimates
of source fluxes may be affected by confusion in the count rate range
below 0.5~cnt~s$^{-1}$, or for fluxes below $7\times
10^{-12}$~erg~s~cm$^{-2}$ (over 3--20~keV), and therefore 
$\sim 20$ of sources entering the catalog at these levels should be
treated with caution.

\begin{figure*}
\includegraphics[width=\textwidth,bb=70 230 570 600,clip]{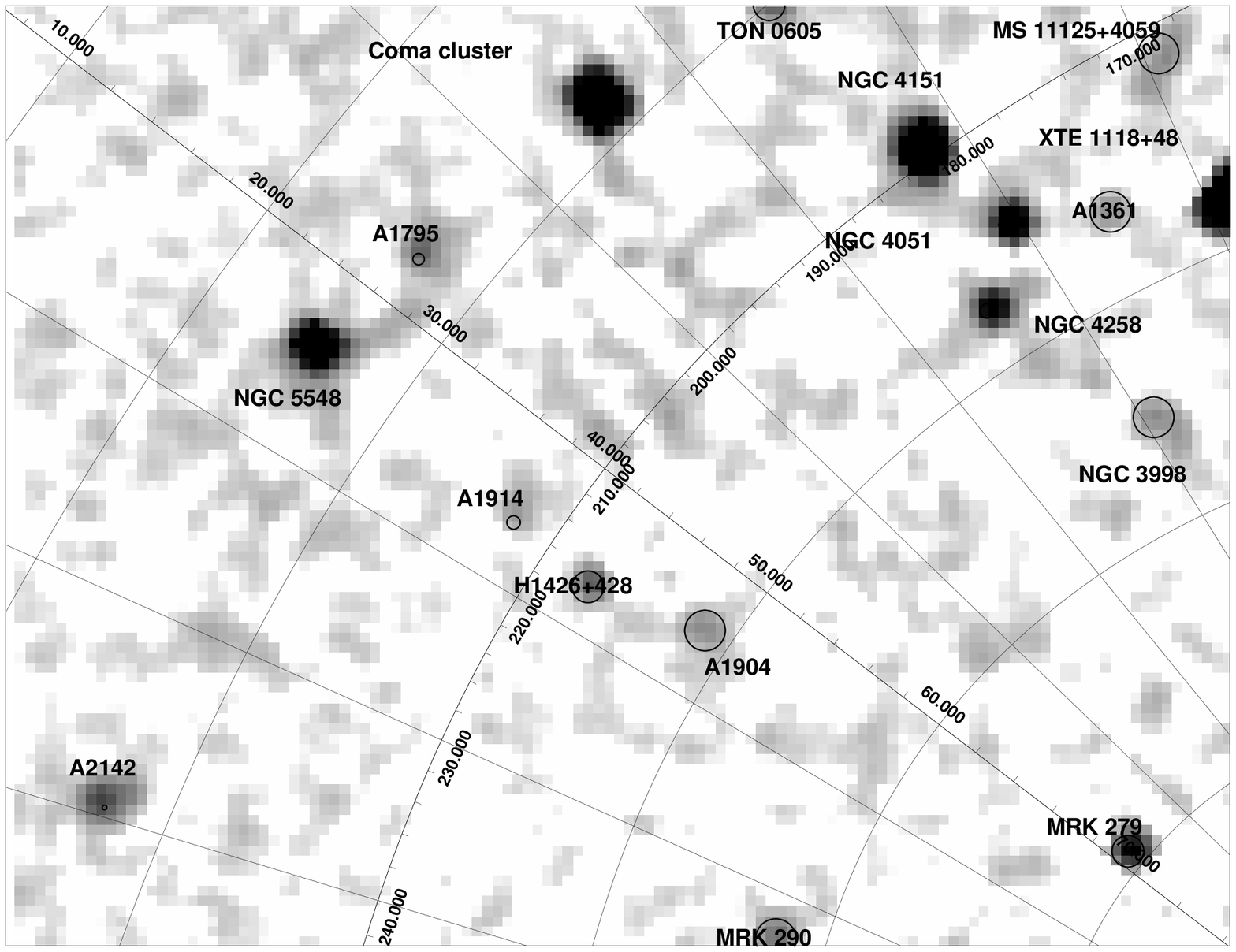}

\caption{Map of the sky around l=80$^\circ$, b=70$^\circ$ in the energy band 
3-20 keV. The gray scale represents the significance of the detection
of X-ray flux from a location in the sky. Circles (very small in the
case of bright sources such as Coma cluster) and labels denote
uncertainty radii of detected sources and their likely
identifications, respectively.
\label{region_map}} 
\end{figure*}

\section{The catalog}
\label{cat}

A total of 294 sources detected (with higher than $4\sigma$ significance) 
in the 3--20~keV energy band at Galactic latitudes $|b|>10^\circ$
comprise the RXTE slew survey catalog, which is presented in
Table~2. The columns of the catalog are described below.

{\it Column (1)} -- source entry name in the catalog. ``XSS'' stands
for ``XTE Slew Survey''. The name provides the source equatorial
coordinates (2000).

{\it Columns (2,3)} -- source Galactic coordinates.

{\it Column (4)} -- localization region radius. In cases where 
the standard procedure of fitting the collimator response to the
source brightness profile is unreliable, a conservative estimate
of the total localization uncertainty is given. Otherwise, the 
quoted value represents 1$\sigma$ statistical uncertainty.

{\it Columns (5,6)} -- count rate and 1$\sigma$ statistical
uncertainty (in cnt s$^{-1}$ per PCU) for the 3--8~keV energy band.  

{\it Columns (7,8)} -- count rate and 1$\sigma$ statistical
uncertainty (in cnt s$^{-1}$ per PCU) for the 8--20~keV energy band. 

{\it Column (9,10)} -- effective slope (photon index) of the source
spectrum and 1$\sigma$ uncertainty. This estimate is based on the
ratio of the 8--20 keV and 3--8 keV count rates. A lower (upper) limit
is given in cases where the upper (lower) band flux measurement has a
signal-to-noise ratio of less than 2.

{\it Column (11)} -- common name of astrophysical object with which
the XSS source is identified. There may be several entries per XSS
source. The identification procedure is described in \S\ref{ident}. 

{\it Columns (12)} -- general astrophysical type of the associated
object. L -- local (source in the Milky Way, LMC or SMC), A --  
active galactic nucleus, C -- cluster of galaxies, G -- non-active
galaxy, O -- other -- afterglow of a gamma-ray burst.

{\it Columns (13)} -- more detailed classification of object. For AGN (A): 
Q -- quasar, RQQ -- radio-quite quasar, RLQ -- radio-loud quasar, BL
-- blasar (BL Lac object or flat-spectrum radio quasar), S1 --  
Seyfert 1 galaxy (types 1, 1.2 and 1.5), NLS1 -- narrow-line Seyfert 1
galaxy, S2 -- Seyfert 2 galaxy (types 1.8, 1.9 and 2), RG -- radio
galaxy, BLRG -- broad-line radio galaxy, NLRG -- 
narrow-line radio galaxy, LLAGN -- low luminosity AGN. AGN 
classification is mostly adopted from NED, otherwise references are
provided. For ``local'' sources (L): XB -- X-ray binary, P -- polar,
IP -- intermediate polar, DN -- dwarf nova of SU, UG or ZC type, NL --
nova-like of VY type, S -- hot star, RS CVn -- RS CVn variable star, SS --
symbiotic star.

{\it Columns (14)} -- redshift (if known) for extragalactic sources.

{\it Columns (15)} -- additional notes or references.

Table \ref{rates_fluxes} provides a set of conversion coefficients
allowing one to estimate the flux of a cataloged source in the
3--8~keV, 8--20~keV and 3--20~keV bands in units of erg~s$^{-1}$~cm$^{-2}$ from
the count rates given in the catalog for a number of values of the
effective spectral slope.
 
\begin{table}
\caption{Conversion factor ($A$) between the RXTE/PCA count rates
($CR$) and the source energy flux ($F$) as a function of the source 
spectral slope ($\Gamma$):
$F$~($10^{-11}$~erg~s$^{-1}$~cm$^{-2})=A\times CR$~(cnt~s$^{-1}$).
\label{rates_fluxes}}
\begin{tabular}{l|l|l|l|l|l|l}
	&       -1&     -0.5&   0&      1&      2&      3\\
\hline
3--8 keV&       0.67&   0.88&   1.00&   1.04&   1.03&   1.02\\
8--20 keV&      2.71&   2.89&   2.87&   2.45&   1.92&   1.51\\
3--20 keV&      2.29&   2.39&   2.27&   1.75&   1.33&   1.12\\
\hline
\end{tabular}
\end{table}

\subsection{Identification of sources}
\label{ident}

We have searched for likely counterparts to the hard X-ray sources
detected in our survey primarily making use of large astronomical databases
including NED, SIMBAD and VizieR. The identification process was
greatly faciliated thanks to the availability of the ROSAT all-sky survey
bright source catalog (RBSC, \cite{rbsc}) of soft X-ray (0.1--2.4~keV)
sources. The ROSAT survey is typically an order of magnitude more
sensitive than our RXTE slew survey for sources with Crab-like spectra
($\Gamma\sim 2$). Therefore, in most cases we could unambiguously
associate the XSS source with a single bright (typically $\ga 0.2$ cnt
s$^{-1}$) and hard (typically $HR1>0.5$) RBSC source located 
within the RXTE error box. Furthemore, for most of such highly probable
associations the astrophysical type is known and other related
information (such as redshift) is available, and that is included in
the catalog. In addition, the ASCA Medium Sensitivity Survey (the GIS catalog,
\cite{gis2}) played a crucial role in deciding about the
identification of a number of sources.

In those cases where we strongly believe that the XSS source
is associated with a particular RBSC source but its nature remains unknown,
we just quote the ROSAT name in the identification column. 
In addition, among the sources of unknown nature in our catalog there
are 6 whose localizations are consistent with a HEAO-1/A1
source (\cite{a1}). Such associations are also indicated in the catalog.

Very hard sources in our catalog such as Seyfert 2 galaxies may
have no obvious counterpart in the RBSC catalog, and even in the ROSAT
all-sky survey faint source catalog, due to a low-energy
photoabsorption cutoff in their X-ray spectrum. We thus carefully 
browsed published samples of AGN, in particular Seyfert 2s, observed
by focusing X-ray telescopes operating in the standard 2--10~keV band
(TARTARUS/ASCA database, \cite{bassani}), searching for possible
associations with our sources.

In addition, we have identified 14 XSS sources (including 3 confused
cases) located in the so-called Zone of Avoidance at
$10^\circ<|b|<20^\circ$ with clusters of galaxies recently discovered
in the CIZA survey (\cite{ciza1}, \cite{ciza2}). 

For completeness, we included in the catalog four sources associated 
with gamma-ray burst afterglows. These detections have become possible
only due to the RXTE observatory targeting these afterglows and
therefore represent an extremely biased statistics of the occurence of
GRB X-ray afterglows on the sky. The effect of cataloged sources being
RXTE targets on the statistics of the catalog is discussed below.

\subsection{Effect of RXTE pointing at cataloged sources}
\label{target}

149 out of the 294 detected sources in our catalog were at least once
the target of RXTE/PCA pointed observations. Although we have not used
the data of the pointed observations themselves, the slew parts of these
observations were included in our analysis. One may therefore
reasonably suspect that some sources have entered the catalog only
because they had been chosen for observations with RXTE. 

In order to check this hypothesis we have re-evaluated the
significance of detection of all cataloged sources by excluding the data
of the slews associated with pointing at the sources. It turns out 
that apart from the GRB afterglows only the sources Mrk 335, Mrk 348,
Ton S180 and NGC 1068 (all are AGN) would not have qualified for a catalog
entry if they had not been pointed at by RXTE. Since this affects only
$\sim 1$\% of the catalog, we conclude that the effect of targeting on the
statistical quality of the catalog is negligible. 

\section{Analysis of the catalog}

Out of the 294 detected sources, 236 (80\%) have been associated
with a single known astrophysical object, another 22 probably result
from superposition of 2 or 3 closely located known
sources. 35 detected sources remain unidentified, although for 12 of
these there is a probable soft X-ray counterpart from the RBSC. Of the
reliably identified single sources, 63 have local origin (among them
are 19 X-ray binaries, 7 polars and 12 intermediate polars), 64 are
clusters of galaxies and 100 are AGN. 

\subsection{Unidentified sources}
\label{unident}

The astrophysical origin of 35 sources in our catalog remains unknown. For 5 of
these detections, the measured 3-20 keV count rate falls below our 
adopted confusion limit of 0.5~cnt~s$^{-1}$ (see \S\ref{flux}), so
some of them may in fact result from superposition of weaker
sources. 29 unidentified sources are newly discovered hard X-ray
sources, while the remaining 6 positionally coincide with previously
known HEAO-1 sources (\cite{a1}). It should be relatively
straighforward to identify in other bands of the electromagnetic
spectrum the 12 XSS sources that have a likely RBSC counterpart,
because the positions of these sources are known to within
1~arcmin. However, the identification of the other sources can be more
problematic. We note that the current localization error ($\sim
30^\prime$) can be significantly reduced by performing slowly scanning
RXTE observations of the sources. 

\begin{figure}
\includegraphics[bb=339 188 574 692,clip,width=\columnwidth]{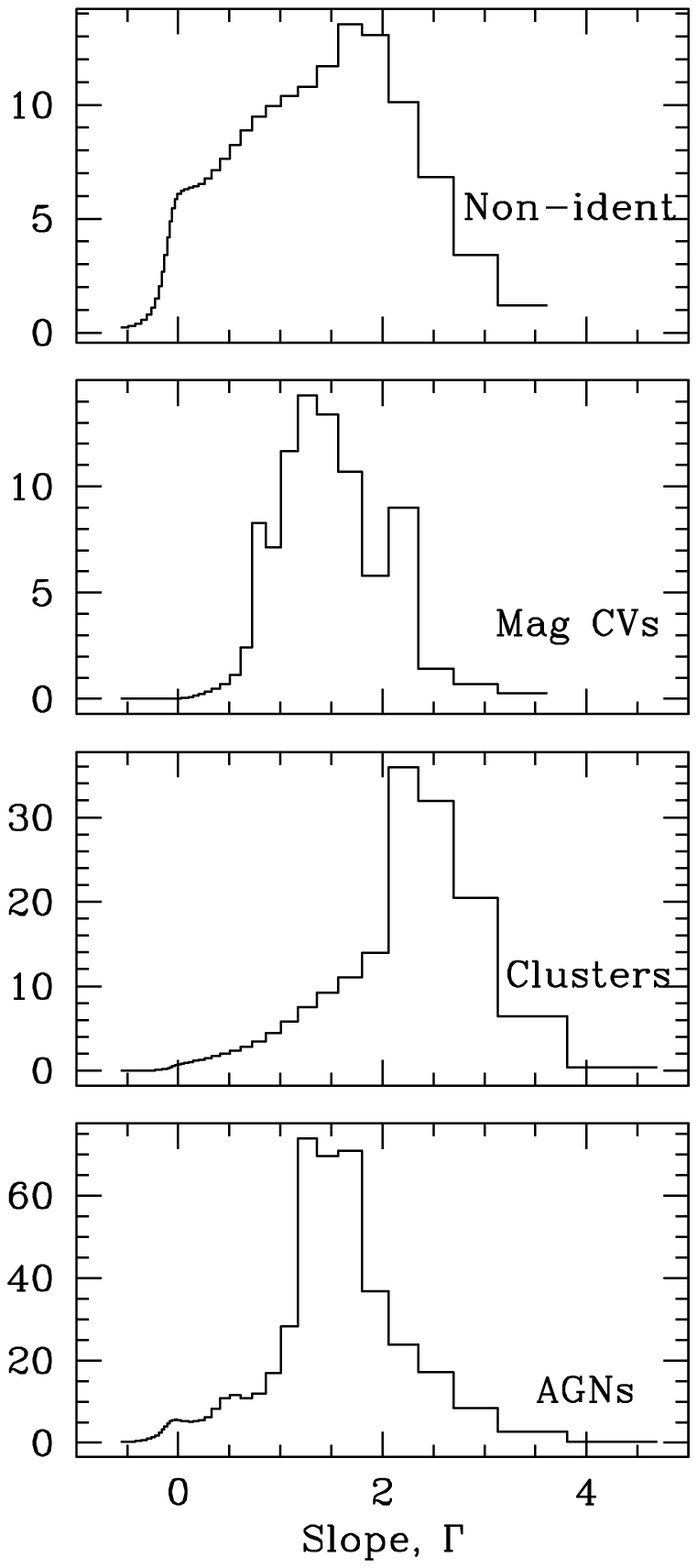}  
\caption{Distribution of spectral slopes of X-ray sources detected in our
survey grouped according to their astrophysical type. The contribution
of each source to the corresponding diagram was computed assuming
that its 8--20~keV/3--8~keV hardness ratio is described by a Gaussian
distribution with the parameters determined by the values and
uncertainty of the measured count rates. 
\label{slopes}}  
\end{figure}

We could get some idea as to the nature of the unidentified sources by
comparing their effective spectral slopes with those of the XSS
sources of known type. This is done in Fig. \ref{slopes}, where we
show the distribution of values of the effective photon index for XSS
sources divided into several major classes, including 
AGN, clusters of galaxies, magnetic cataclysmic
variables (CVs, including polars and intermediate polars) and
unidentified sources. In plotting these diagrams, each source was
ascribed a distribution of hardness ratios (8--20~keV/3--8~keV) in
accordance with the measured count rate uncertainties.  

The resulting distribution of $\Gamma$ values is apparently 
broad for the sample of unidentified sources, although substantial
part of this scatter results from the statistical uncertainty in
measuring the count rates. However, the bulk of the unidentified sources
are evidently harder than typical clusters of galaxies, being similar in
hardness ($\Gamma\la 2$) to AGN and magnetic CVs.

It is interesting that 31 of 35 unidentified sources are located in
the southern hemisphere (at $\delta <0$), which probably reflects the
fact that the southern sky is relatively less studied in the optical
and other bands. We can use this fact to show that magnetic CVs are
unlikely to constitute the majority of our unidentified sources. Indeed, 
8 and 11 of the magnetic CVs in our catalog are located at $\delta>0$ and
$\delta<0$, respectively. Since these objects are located in space
within $\sim 500$~pc from us, their distribution in the sky is
expected to be random to a first approximation. This statement relies
on the fact that for a given source luminosity our survey samples
similar volumes in the northern and southern hemispheres (in 1 to 1.1
proportion). For comparison, the catalog of CVs by Ritter et al. 2003,
which contains 107 polars and intermediate polars and among them all
of our 19 magnetic CVs, does not demonstrate any significant
north/south asymmetry either: 49 and 58 sources are located at $\delta>0$ at
and $\delta<0$, respectively. If many of our unidentified
sources were magnetic CVs, the distribution of all detected sources of
this type would become significantly asymmetric with respect to the
equatorial plane, which would be very difficult to explain.

Our catalog also includes 44 local sources of types other than
magnetic CVs. However, we do not expect that more than a few 
objects of these types may be present among our unidentified
sources. Indeed, X-ray binaries are very rare and bright X-ray sources,
while the other sources (such as hot stars) are also rare and usually
have soft spectra ($\Gamma>2$), as evidenced by our catalog.

We conclude that most of the unidentified sources are likely AGN,
including those absorbed at soft X-rays (such as Seyfert 2 galaxies). The AGN
content of the RXTE/PCA all-sky slew survey will be discussed in detail in a
separate paper.

\subsection{Number-flux functions}
\label{nflux}

The high completeness ($\sim 90$\%) of the identification of detected
sources allows us to estimate number-flux functions for extragalactic
sources. Furthemore, thanks to the record sensitivity achieved by the
current survey at 8--20~keV, we obtain a unique opportunity to
construct a $\log N-\log S$ distribution in this hard X-ray band. 

\begin{figure*}
\hbox{
\includegraphics[width=\columnwidth]{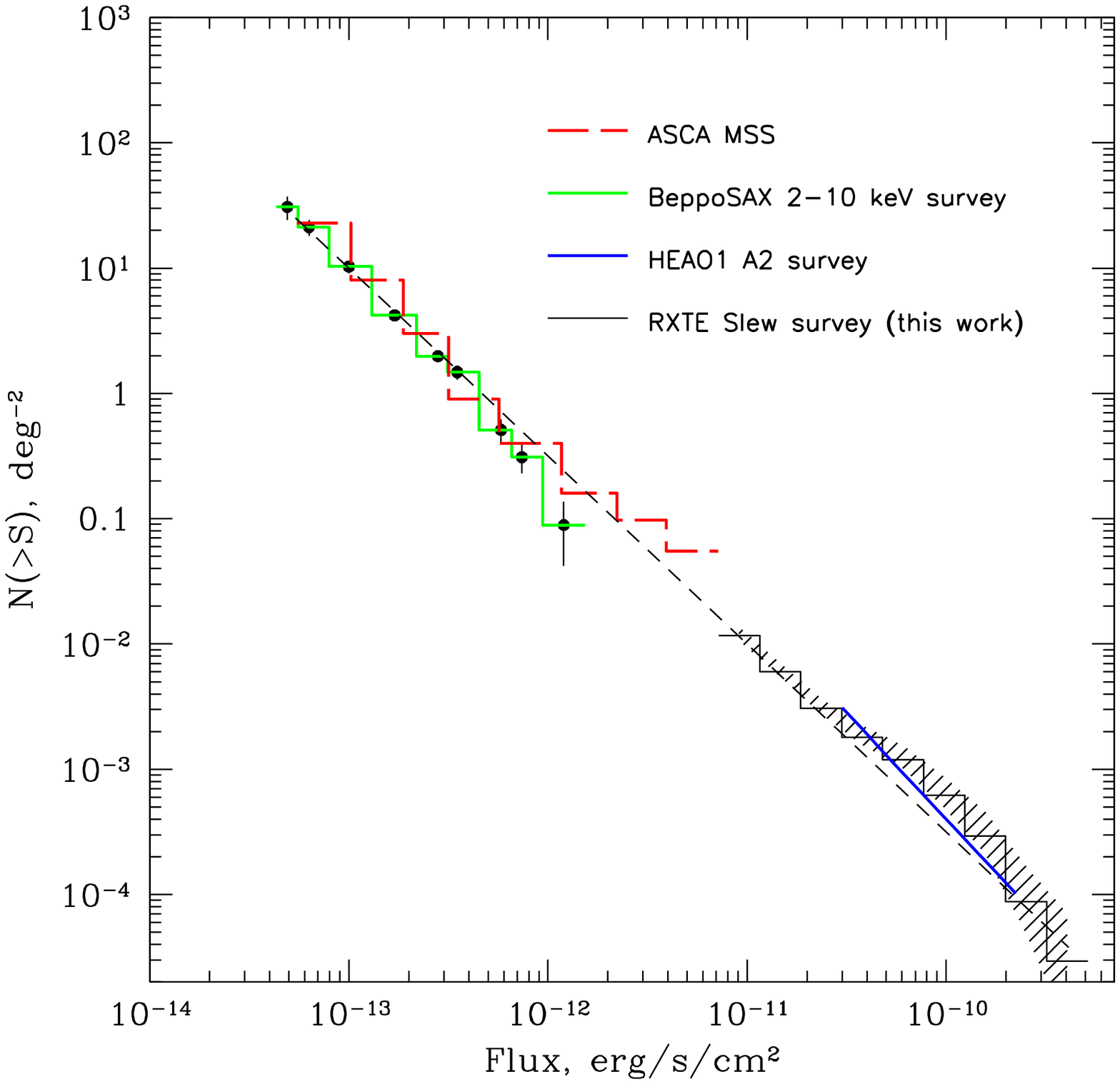}
\includegraphics[width=\columnwidth]{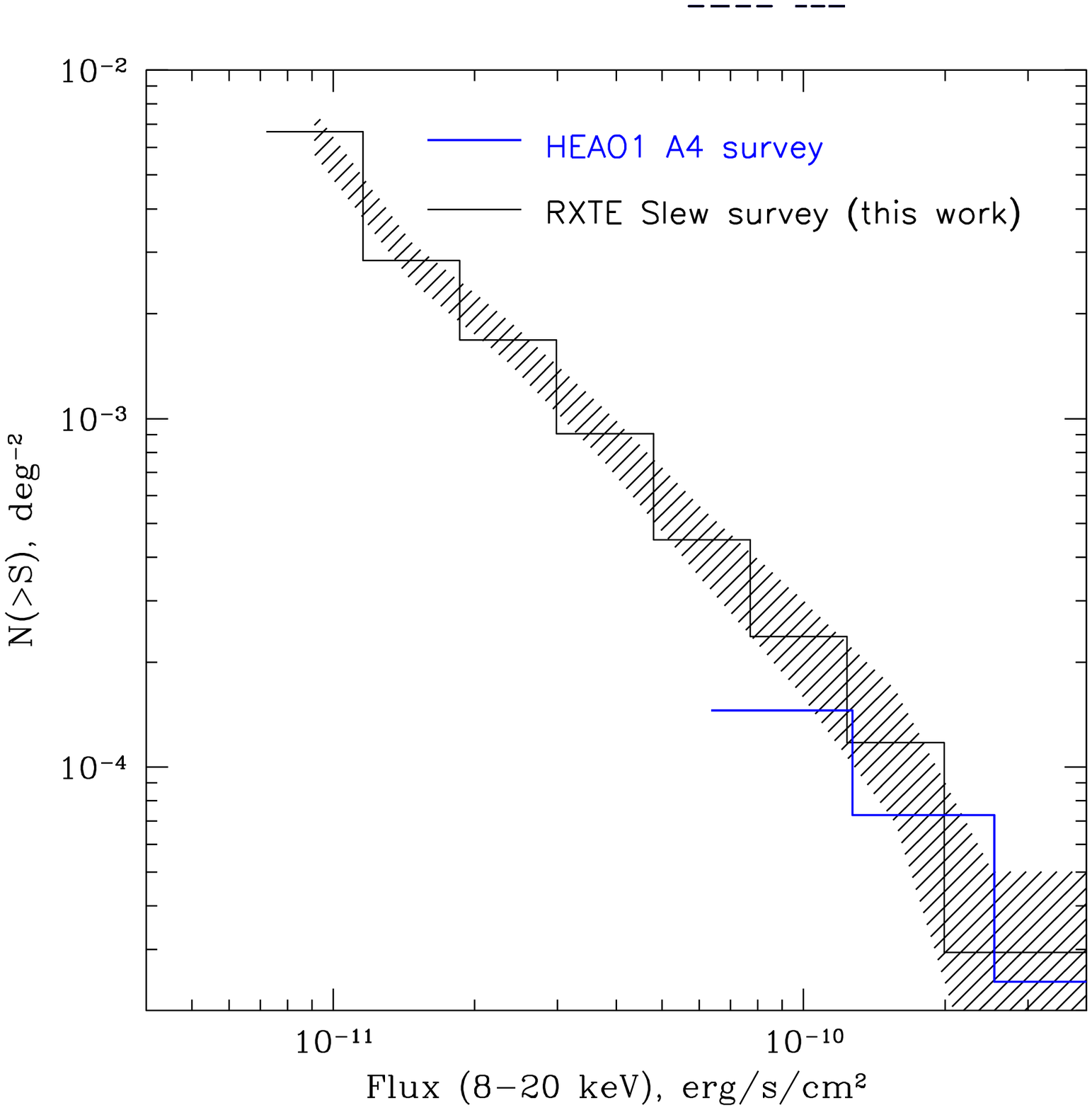}
}
\caption{{\bf Left}: Cumulative $\log N-\log S$ function in the
2--10~keV band obtained from the RXTE all-sky slew survey. To obtain
this plot, the count rates measured at 3--8~keV were converted to
2--10~keV fluxes assuming a power law spectrum with a photon index
$\Gamma=2$. The dashed area represents the estimated statistical
uncertainty. Also shown are a number of previous measurements 
with: HEAO-1/A2 (Piccinotti et. 1982), ASCA (Ueda et al. 1999) and
BeppoSAX (Giommi et al. 2000). The short-dashed line is our best
fit with the slope $\alpha=1.5$ to the RXTE data extrapolated down to
the fluxes sampled by the BeppoSAX and ASCA surveys. {\bf Right}: Cumulative
$\log N-\log S$ function in the 8-20 keV band obtained from the RXTE
slew survey. Also shown is the result of the HEAO-1/A4 experiment
(Levine et al. 1984) recalculated from the 13--25~keV band to
8--20~keV assuming a $\Gamma=2$ spectrum.  
\label{lognlogs_soft}}
\end{figure*}

Fig. \ref{lognlogs_soft} shows the cumulative flux distributions
of the detected extragalactic sources in the 2--10~keV and 8--20~keV
energy bands, estimated as follows. We selected from the general
catalog (Table 2) two samples of sources 1) those having $S/N>4$ in
the 3--8~keV band and 2) those meeting the same condition in the
8--20~keV band. Excluding sources of confirmed local origin, we are
left with 209 and 92 sources for the soft and hard bands,
respectively. These samples include unidentified sources (27 and 7,
respectively), although the results remain essentially unchanged if we
exclude these cases from consideration (8 unidentified sources have $S/N<4$
in the soft energy band and 28 sources -- in the hard energy band). 
As was explained in
\S\ref{map}, although the general catalog comprises sources originally
detected in the broad 3--20~keV band, the source samples subsequently
selected from it in the 3--8~keV and 8--20~keV subbands are expected
to be highly complete. 

The $\log N-\log S$ curves were calculated using the known sky
coverage as a function of the survey sensitivity in the 3--8~keV and
8--20~keV bands (Fig. \ref{exposure_area}) The
obtained distributions were converted from count rates to flux 
units assuming a power-law spectrum of $\Gamma=2$. In addition, to
faciliate comparison of our results with those of previous experiments
we have made a change from the 3--8~keV band used in the catalog to
the standard 2--10~keV band. Finally, we truncated the $\log N-\log S$
distributions at the lowest fluxes according to the confusion
limit formulated in \S\ref{flux}, namely at $6\times
10^{-12}$~erg~s~cm$^{-2}$ and $4\times 10^{-12}$~erg~s~cm$^{-2}$ in
the 2--10~keV and 8--20~keV, respectively. These values result 
from the fits to the number-flux functions at higher fluxes, presented
below. As a result, the final flux distributions for the soft and hard bands
presented in Fig. \ref{lognlogs_soft} are based on 197 and 90 sources,
respectively.

Assuming the usual power-law form for the number-flux function
$$
N(>S)=K S^{-\alpha},
$$
we can estimate the parameters of this function using the maximum
likelihood method (e.g \cite{crawford70}). We obtain the following
best fits to the observed distributions in the flux ranges
$6\times10^{-12} -  6\times 10^{-10}$~erg~s~cm$^{-2}$ (3--8~keV) and
$4\times10^{-12} -  4\times 10^{-10}$~erg~s~cm$^{-2}$ (8--20~keV):
$$
N(>S_{11})=(9\pm1)\times 10^{-3} S_{11}^{-1.34\pm0.13}
\,\,{\rm deg}^{-2}\,\,{\rm (2-10~keV)} 
$$
and
$$
N(>S_{11})=(5.8\pm0.6)\times 10^{-3} S_{11}^{-1.52\pm0.18}
\,\,{\rm deg}^{-2}\,\,{\rm (8-20~keV)},
$$
where $S_{11}$ is the source flux in units of
$10^{-11}$~erg~s$^{-1}$~cm$^{-2}$, the uncertainties here and later
being $1\sigma$.
 
If we fix the slope of the number-flux function at the value $\alpha=1.5$,
we obtain normalization constants $K=(1.0\pm0.1)\times 10^{-2}$ and
$K=(5.6\pm0.6)\times 10^{-3}$~deg$^{-2}$ for the soft and hard energy
bands, respectively.

It is interesting to compare the above normalization of the 2--10~keV
cumulative source counts with the values inferred from previous sky
surveys, carried out with Ariel 5 (\cite{ariel5}), UHURU (\cite{4u}), 
HEAO-1/A2 (\cite{a2}) and BeppoSAX (\cite{giommi2000}). Adopting
conversion factors of 1~cnt~s$^{-1}$ (Ariel 5) = 
$6.2\times10^{-11}$~erg~s$^{-1}$~cm$^{-2}$, 1 cnt~s$^{-1}$ (UHURU) =
$2.4\times10^{-11}$~erg~s$^{-1}$~cm$^{-2}$ and 1 cnt~s$^{-1}$  
(HEAO-1/A2) = $2.4\times10^{-11}$~erg~s$^{-1}$~cm$^{-2}$ for the 2--10 keV
energy band, we find the following coefficients (in the
units given above) of the cumulative distribution function:
$K_{\rm Ariel\,5}= (1.5\pm 0.4)\times 10^{-2}$, $K_{\rm UHURU}= 
(1.8\pm 0.2)\times 10^{-2}$, $N_{\rm HEAO-1/A2}=(1.24\pm0.14)\times 10^{-2}$, 
$K_{\rm BeppoSAX}=8.9\times 10^{-3}$.

It can be seen that the Ariel 5, UHURU and HEAO-1 normalizations are
slightly higher than ours. This difference probably results from the
fact that our estimate of the density of extragalactic sources is
largely based on low flux ($\la 2\times
10^{-11}$~erg~s$^{-1}$~cm$^{-2}$) sources that were inaccessible to  
those experiments. Our result is therefore less affected
by the substantially inhomogeneous distribution of X-ray sources in
the local extragalactic cell ($\la 100$~Mpc in size),
because we efficiently sample the local Universe outside this volume
as well. We note that the source counts measured with RXTE at flux levels above
a few $\times 10^{-11}$~erg~s$^{-1}$~cm$^{-2}$, where the 
contribution of the local density inhomogeneity becomes important, 
lie above our power-law fit and are in fact consistent with
the results of the early all-sky surveys mentioned above. On the other
hand, the extrapolation of our $\log N-\log S$ curve to fluxes below
the limit of the current survey of $6\times 10^{-12}$
erg~s$^{-1}$~cm$^{-2}$ (2--10~keV) is in excellent agreement with the flux
distributions derived from the BeppoSAX and ASCA medium-sensitivity
surveys (see Fig. \ref{lognlogs_soft}).
 
\subsection{Summary}
\label{sum}

We presented the serendipitous hard X-ray (3--20~keV)
survey of the sky at high Galactic latitude $|b|>10^\circ$ based on
RXTE/PCA slew observations in 1996--2002. At photon energies above
10~keV, the current survey surpasses by an order of magnitude in depth
the previously best HEAO-1/A4 survey, achieving similar sensitivity
($\sim 2\times 10^{-11}$~erg~s$^{-1}$~cm$^{-2}$) to the
HEAO-1/A1 experiment in the standard X-ray band (2--10~keV) for the most
of the sky. At the same time, $\sim 20$\% of the sky is now covered at
flux levels below $10^{-11}$~erg~s$^{-1}$~cm$^{-2}$ (2--10~keV), which  
has allowed us to draw the $\log N-\log S$ diagram for extragalactic
X-ray sources down to $6\times 10^{-12}$~erg~s$^{-1}$~cm$^{-2}$ and thus to
fill the previously existing gap toward the cumulative source counts measured
at lower fluxes by medium sensitivity surveys.  Moreover, we have  
for the first time estimated the distribution of source fluxes at 8--20~keV
in the $4\times10^{-12}$--$4\times 10^{-10}$~erg~s$^{-1}$~cm$^{-2}$ range. 

The compiled catalog comprises 294 detected sources and provides for
each of them the estimated flux and effective spectral slope in the 
3--20~keV range. The identification of the sources is already highly 
complete (88\%), so the catalog can be efficiently used for statistical
studies of different classes of sources. However, particularly
interesting scientific content might still be hidden in the sample of 35
unidentified sources. A large fraction of these are probably AGN,
including a number of highly obscured ones which are very difficult to
discover in the optical or soft X-ray bands. To establish their
origin, a dedicated identification program needs to be carried out.  Scanning
RXTE observations could improve the localizations of those
unidentified sources for which we failed to find a counterpart in the ROSAT
all-sky survey. 

The statistical properties of the AGN detected in the RXTE all-sky
slew survey will be discussed in a separate paper (Sazonov et al., in
preparation). 

\begin{acknowledgements}
We thank Harald Ebeling and Christopher Mullis for providing us the
information about a set of CIZA clusters before publication.
This research has made use of data obtained through the High Energy
Astrophysics Science Archive Research Center Online Service,
provided by the NASA/Goddard Space Flight Center.
\end{acknowledgements}

\clearpage

\begin{figure*}
\includegraphics[width=\textwidth]{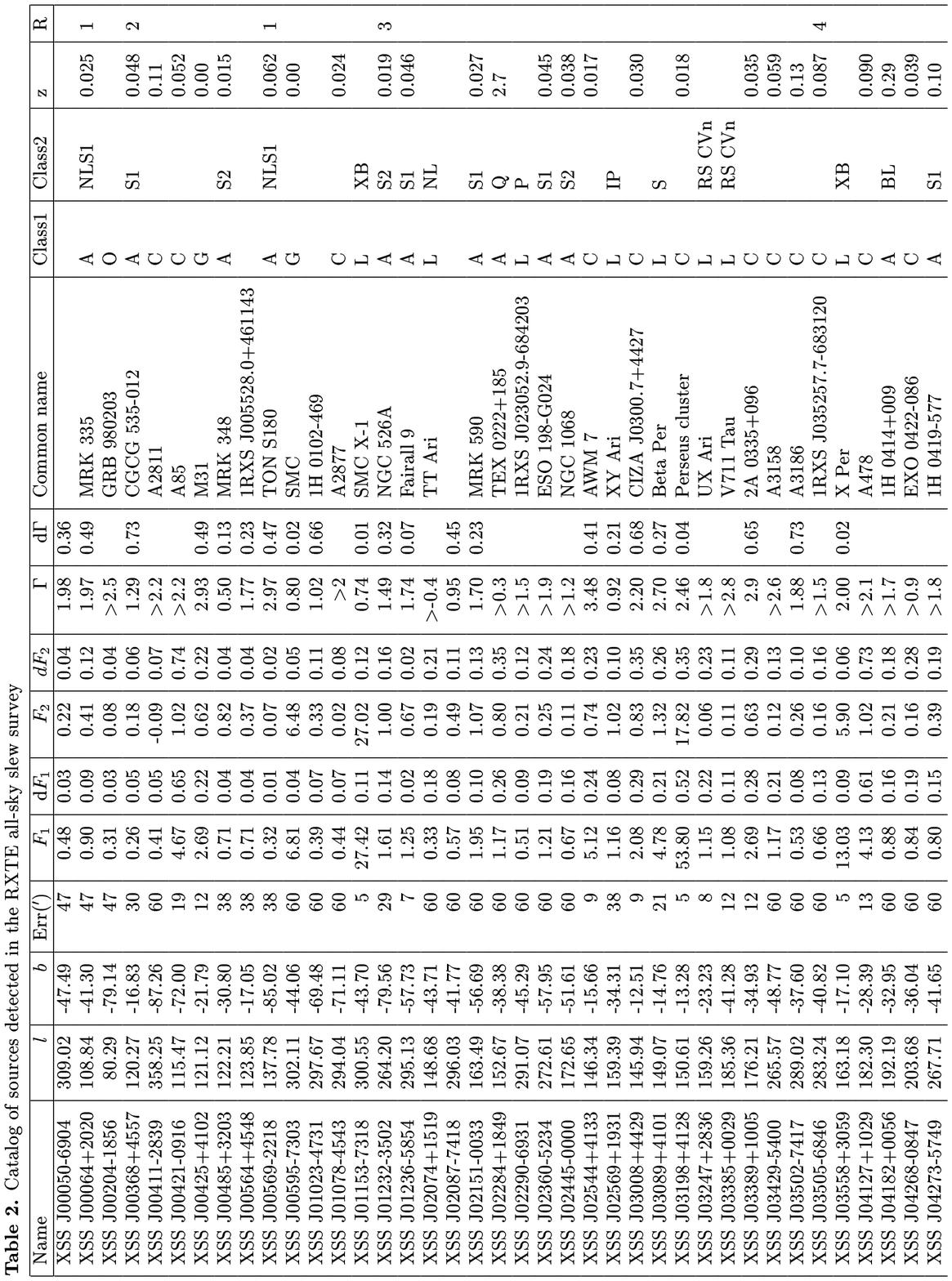}
\end{figure*}

\begin{figure*}
\includegraphics[width=\textwidth]{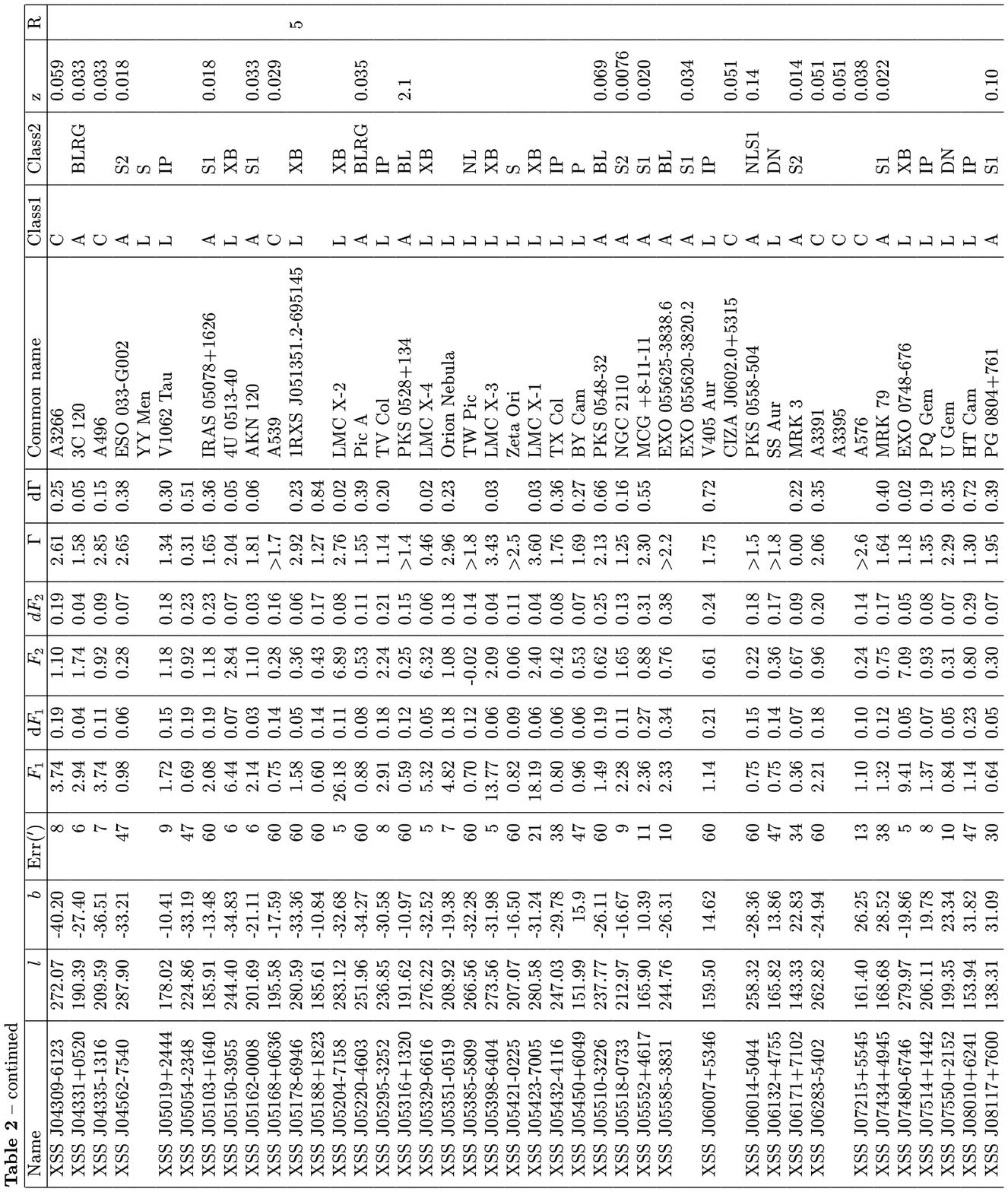}
\end{figure*}

\begin{figure*}
\includegraphics[width=\textwidth]{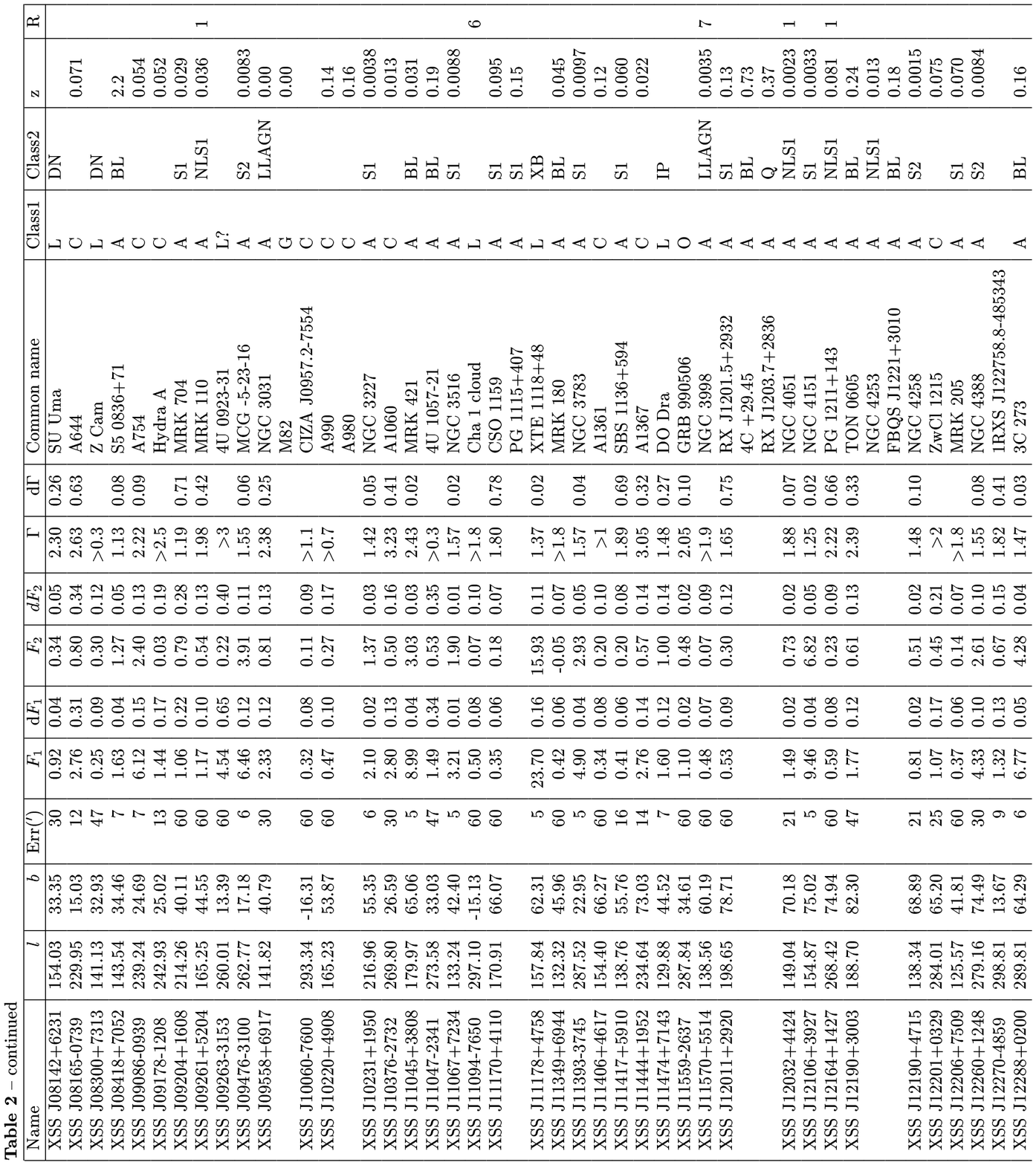}
\end{figure*}

\begin{figure*}
\includegraphics[width=\textwidth]{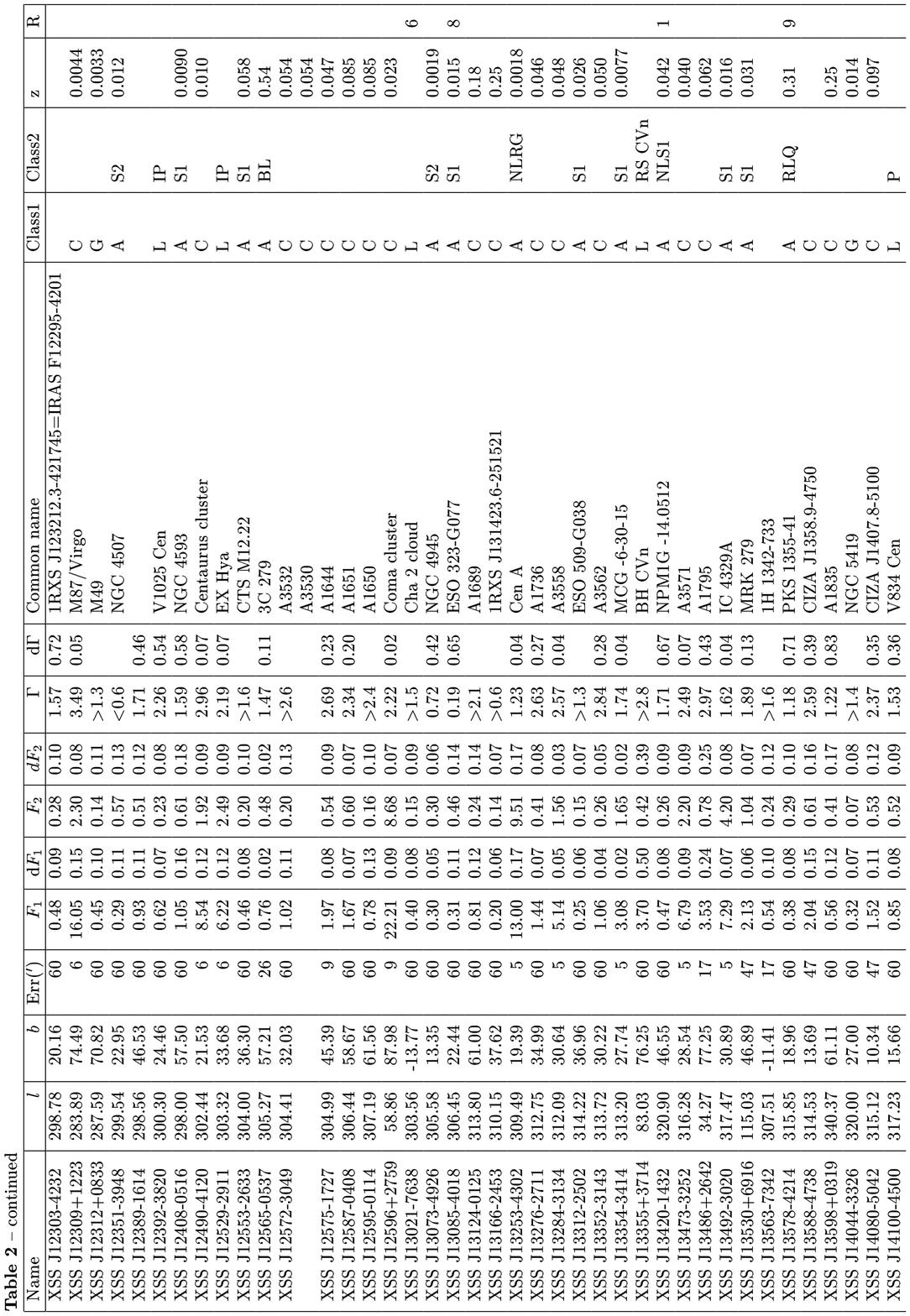}
\end{figure*}

\begin{figure*}
\includegraphics[width=\textwidth]{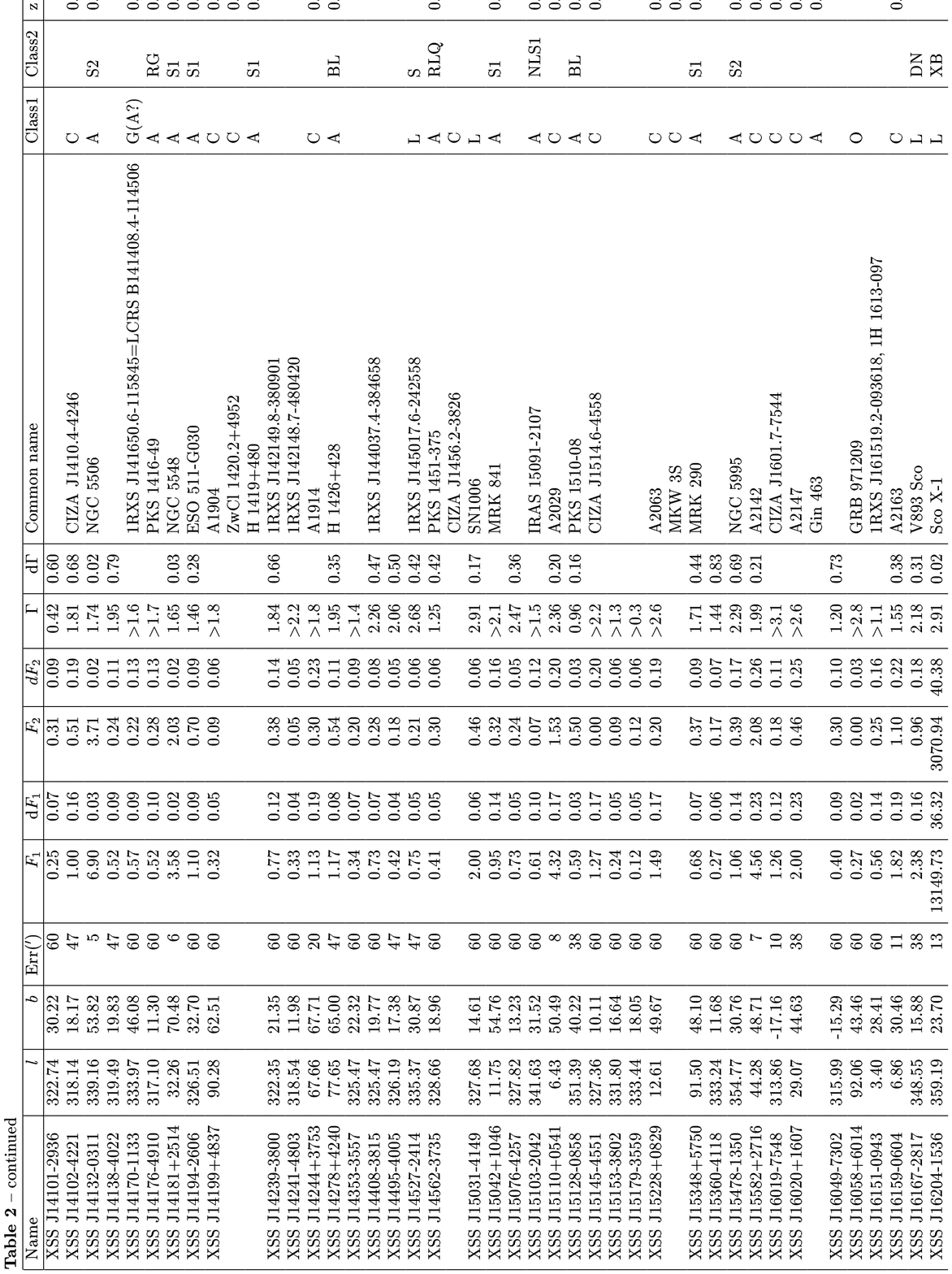}
\end{figure*}

\begin{figure*}
\includegraphics[width=\textwidth]{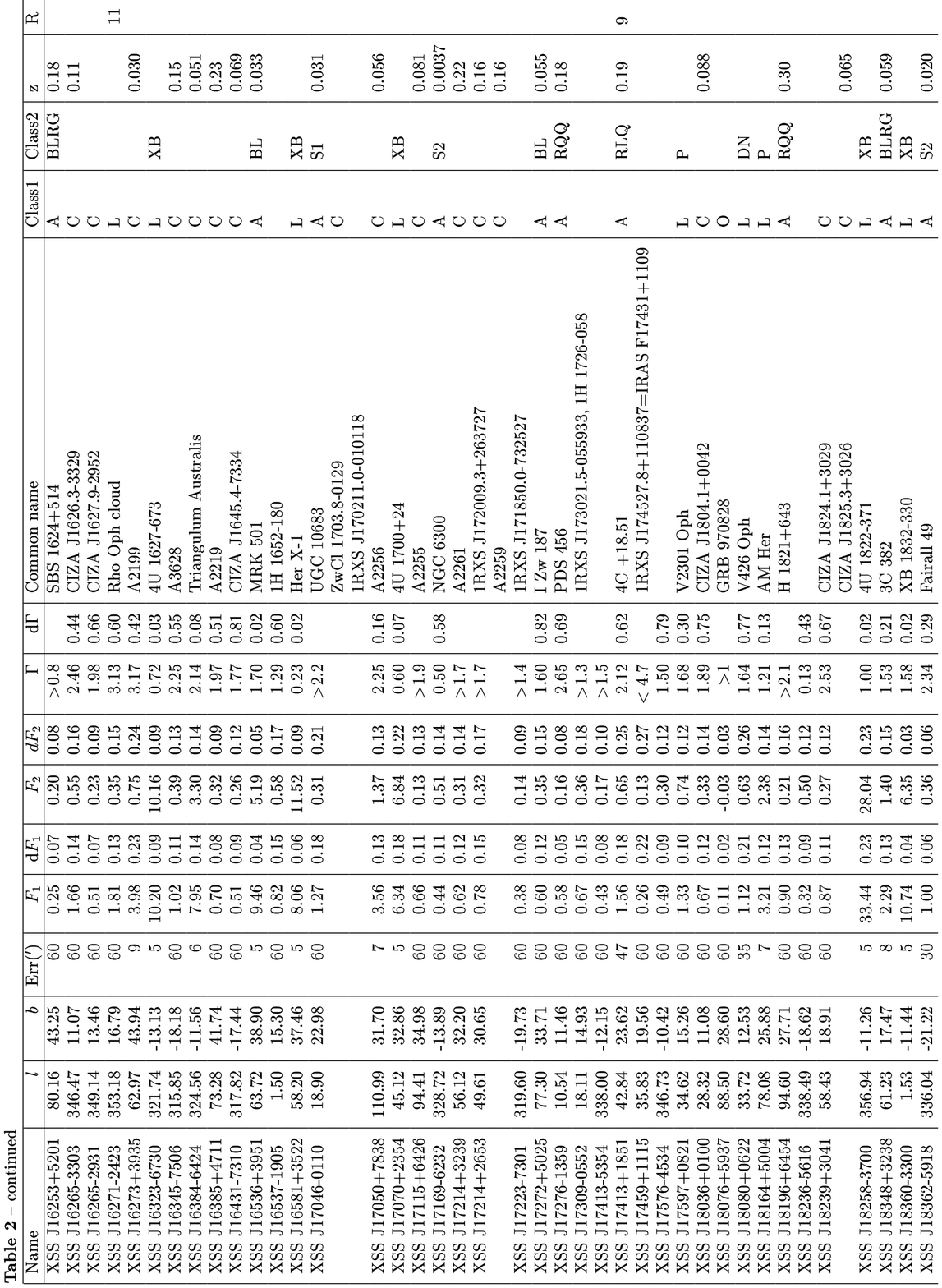}
\end{figure*}

\begin{figure*}
\includegraphics[width=\textwidth]{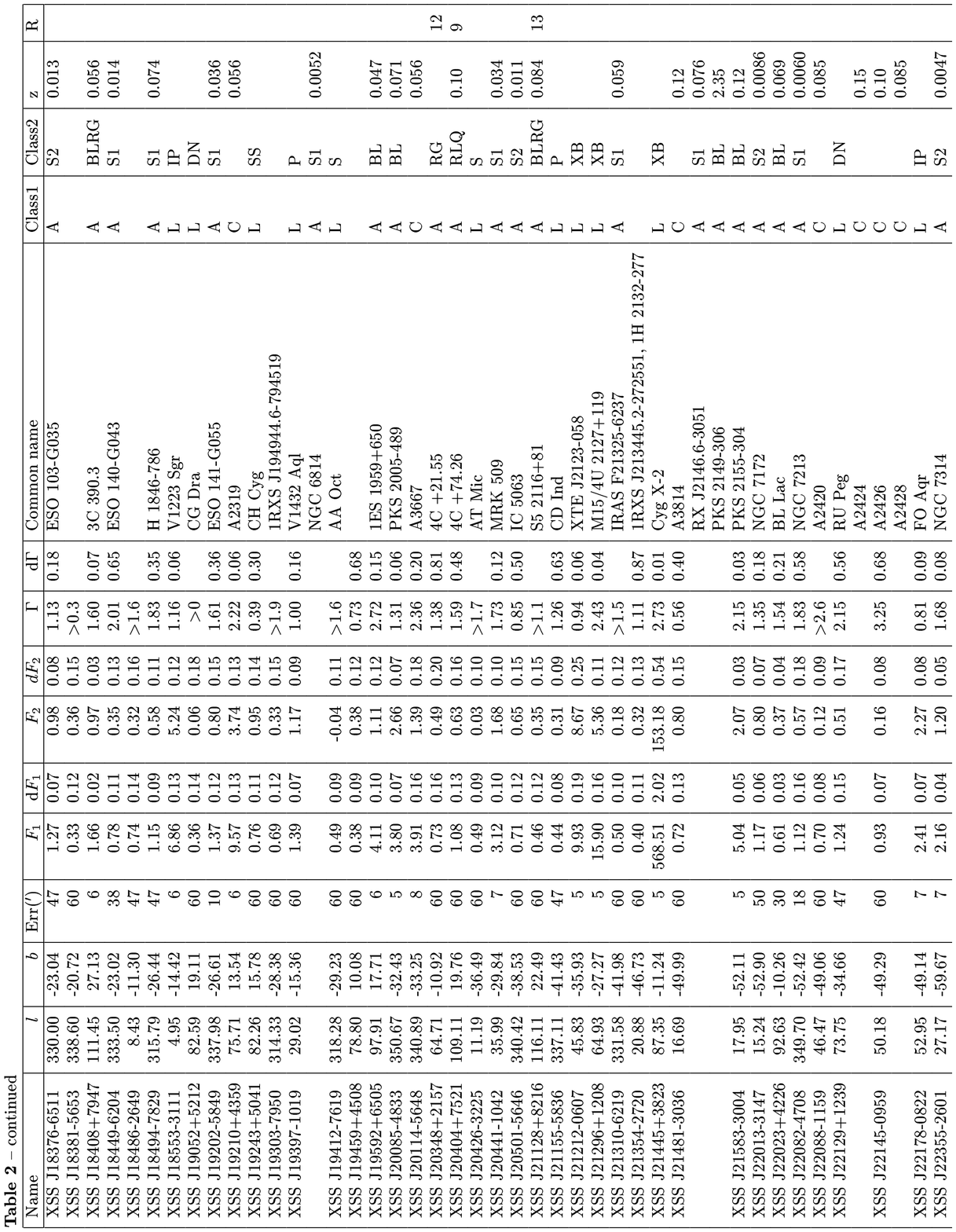}
\end{figure*}

\begin{figure*}
\includegraphics[width=\textwidth]{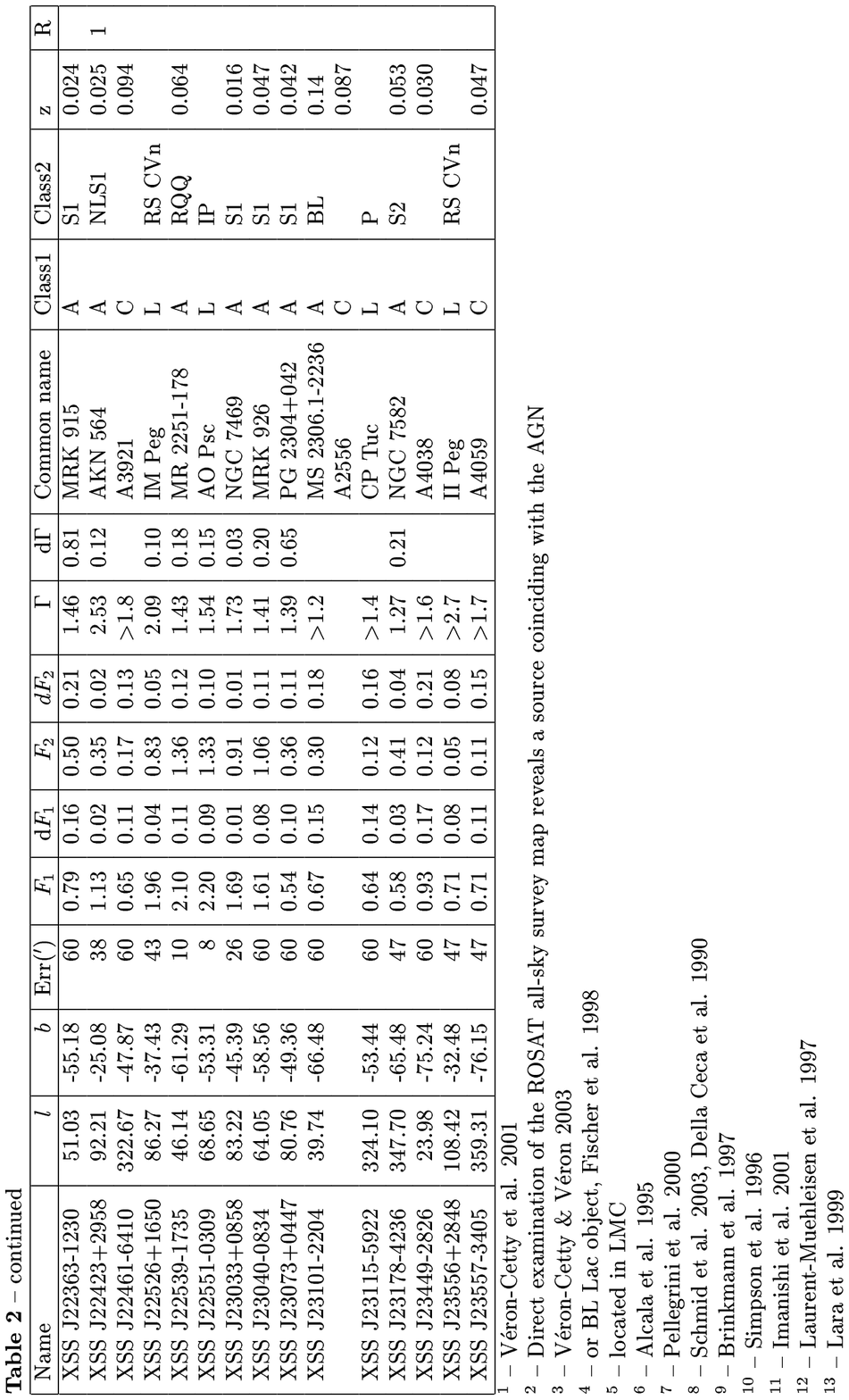}
\end{figure*}

\end{document}